\documentstyle[aps,psfig]{revtex}
\begin{document}

\begin{center}
{\Large\bf Contribution of the massive photon decay channel to
neutrino cooling of neutron stars}\\[1cm] { D.N.
Voskresensky\footnote{ e-mail: voskre@rzri6f.gsi.de}}\\
{\sl Moscow
Institute for Physics and Engineering, Kashirskoe shosse 31,
115409 Moscow, Russia\\ GSI, Planckstr. 1, D-64291, Darmstadt,
Germany}\\[3mm] {E.E. Kolomeitsev\footnote{ e-mail:
kolomei@tpri6f.gsi.de}, B. K\"ampfer}\\ {\sl Institut f\"ur Kern
und Hadronenphysik, Forschungszentrum Rossendorf e.~V.,\\ PF
510119, D-01314 Dresden, Germany\\[2mm] Institut f\"ur Theoretische
Physik, TU Dresden,\\ Mommsenstr. 13, D-01062 Dresden, Germany}
\end{center}

\begin{abstract}
We consider massive photon decay reactions via intermediate states of
electron--electron-holes and proton--proton-holes
into neutrino--antineutrino pairs in the course of neutron
star cooling. These reactions may become
operative in hot neutron stars in the region of proton pairing
where the photon due to the Higgs--Meissner effect
acquires an effective mass $m_{\gamma}$ that is small compared to
the corresponding plasma frequency.
The contribution of these reactions to neutrino emissivity
is calculated;
it varies with the temperature and the photon mass
as $T^{3/2}m_{\gamma}^{7/2}{\displaystyle  e}^{-m_{\gamma}/T}$ for
$T< m_{\gamma}$. Estimates show that these processes
appear as extra efficient cooling channels of neutron stars at
temperatures $T\simeq (10^9-10^{10})$~K.
\end{abstract}

\noindent
PACS number(s): 21.65.+f, 95.85.R, 97.60.I\\
key words: nuclear matter, neutron stars, neutrino cooling

\newpage

\section{Introduction}

The {\small EINSTEIN}, {\small EXOSAT} and {\small ROSAT}
observatories measured surface temperatures of certain neutron
stars and put upper limits on the surface temperatures of others (
cf. Ref.~\cite{steuk} and further references therein). Data on the
supernova remnants in 3C58, the Crab, and RCW103 indicate rather
slow cooling, while the data for Vela, PSR~2334+61, PSR~0656+14,
and Geminga point to significantly more rapid cooling. In the
so-called standard scenario of neutron star cooling, the most
important channel up to temperatures $T\leq (10^{8} - 10^{9})K$
corresponds to the modified URCA process $n \, n
\rightarrow n \, p\, e\, \bar \nu$.  Rough estimates of its
emissivity were first made in Ref.~\cite{BW}. Friman and Maxwell
in Ref.~\cite{fm} recalculated the emissivity of this process in a
model, in which the nucleon--nucleon interaction is treated with
the help of slightly modified free one-pion exchange. Their result
for the emissivity, $\varepsilon _{\nu}^{FM}$, proved to be an
order of magnitude higher than previously obtained. The value
$\varepsilon
_{\nu}^{FM}$ was used in various computer simulations resulting in
the standard cooling scenario; see Ref.~\cite{T} for example.
Subsequent work~\cite{pion,sv,migrep} took in-medium effects into
account
 in $NN$-interaction, showing that the emissivity of
the modified URCA process depends heavily on neutron star mass.
For stars of more than one solar mass, the resulting
emissivities turned out to be substantially higher
than the values given by  $\varepsilon _{\nu}^{FM}$.

These and other
in-medium effects were recently
incorporated in the computer code in Ref.\cite{SVSWW} leading to a
new scenario of neutron star cooling.  For low-mass stars
numerical results of
the new and standard scenarios more or less coincide.
In the present work, we continue to look for
enhanced reaction channels. To demonstrate the efficiency of new
reaction channels, we compare the results with
the emissivity $\varepsilon _{\nu}^{FM}$, which
dominates cooling in the standard scenario over
temperature range under consideration.

Besides the modified URCA process, the standard
scenario numerical codes also include
 neutron and proton bremsstrahlung processes
$n\, n\rightarrow n\, n \nu  \bar{\nu}$ and $n\, p\rightarrow n\, p
\nu  \bar{\nu}$, which in all models lead to a somewhat smaller
contribution to the emissivity than the modified URCA process
\cite{FSB,fm,pion,sv}. Also included are processes that contribute
to emissivity in the neutron star crust. These are plasmon decay
$\gamma_{pl}\rightarrow \nu\, \bar{\nu}$ \cite{arw,SB},
electron bremsstrahlung on nuclei $e\, A\rightarrow e\, A\nu
\bar{\nu}$ \cite{P,rf,SB}, electron--positron annihilation $e\,
e^{+}\rightarrow \nu \bar{\nu}$~\cite{cm,cs}, and photon absorption
by electrons $\gamma e\rightarrow e\, \nu
\bar{\nu}$~\cite{Rit,cs,PBS}. Numerical simulations show that the
latter two processes contribute only negligibly to the crust
neutrino emissivity at the temperatures under discussion in this
paper and they always contribute negligibly to the full neutron
star's emissivity; see Fig.~7 of Ref.~\cite{SB}.

When the temperature decreases, it is energetically favorable for
neutrons to pair in the neutron star interior and  inner crust
and for the protons to pair in the star's interior.
In a system with nucleon pairing the
emissivity of the modified URCA process is suppressed by a factor
$\exp(-(\Delta_n+\Delta_p)/T)$ \cite{fm},
where $\Delta_n$ and $\Delta_p$ are the
respective
neutron and proton gaps, defined by
$\Delta_i(T)=\Delta_i(0) \, (T_{c,i}-T) \, T_{c,i}^{-1} \, \Theta(T_{c,i}-T)$
(here $\Theta(x)$ is the Heaviside step function, $i=\{p,n\}$, and $T_{c,i}$
is the corresponding critical temperature for nucleon pairing).
At temperatures $T\ll T_{c,p}, T_{c,n}$
the process becomes marginal. Nevertheless,
this star's interior process still
dominates those of crust
cooling up to temperatures $T\sim (10^{8} - 10^{9})$~K, depending on the
values of the gaps; see Fig.~7 of Ref.~\cite{SB}.
For $T\leq (1 - 3)\cdot 10^{8}$~K cooling in the standard scenario
is largely dominated by the photon emission from the neutron star surface.

In the present work we look for more efficient cooling processes
at $T< T_{c,p}, T_{c,n}$.
We analyze photon decay into neutrino--antineutrino pairs.
The related processes  $\gamma e\rightarrow e\, \nu \bar{\nu}$
and $\gamma p\rightarrow p\, \nu \bar{\nu}$
turn out to be suppressed by
several orders of magnitude compared to those under discussion,
due to the lack of free final states in
degenerate fermionic systems, and are therefore not considered here.
The contribution of photon decay
via electron--electron-hole intermediate states
for the case of a normal electron plasma
in white dwarfs and neutron star crusts has been calculated
by several authors (see Ref.~\cite{arw} for further references).
In an ultrarelativistic
electron plasma, a photon acquires an effective in-medium
plasmon dispersion law with a gap equal to
the electron plasma frequency
$\omega _{pl}\simeq 2\,e\,\mu_e /\sqrt{3\pi}$,
where $e$ is the electron charge and $\mu_e$ denotes the electron chemical
potential (we employ units with $\hbar = c = 1$). Therefore, the
contribution to the emissivity of the cited process is
suppressed by a factor $\exp(-\omega _{pl}/T)$.
Nevertheless, in white dwarfs and neutron star crusts, the
electron density is not too high, and the process is still effective.
In neutron star interiors, the electron density $\rho_e$
is equal to the proton density $\rho _p$ by virtue
electrical neutrality, and along with $\beta$ stability
one obtains a relation for the total density
\begin{equation}  \label{3}
\rho_e=\rho _p\simeq\, 0.016\,\rho_0\,
\left(\frac{\rho }{\rho_0  }\right)^2,
\end{equation}
where $\rho_0  \simeq\, 0.17$~fm$^{-3}$ denotes the
nuclear saturation
density, and we use the values
 of the neutron and proton Fermi momenta~\cite{fm},
$p_{Fn}\simeq\,340(\rho /\rho_0)^{1/3}$~MeV and
$p_{Fp}=\mu_e \simeq\,85(\rho /\rho_0  )^{2/3}$~MeV.
Thus, at typical densities for neutron star interiors
$\rho \stackrel{>}{\sim} \rho_0$,
the value of the electron plasma frequency is high, e.g.,
$\omega _{pl}(\rho_0)\approx $ 4.7 MeV for $\rho \simeq \rho_0$,
and at temperatures
$T<T_{c,n}, T_{c,p}< \omega _{pl}$ the process
$\gamma_{\omega_{pl}} \to  e \, e^{-1} \to \nu \bar \nu$, where
the superscript $-1$ denotes the hole,
is strongly suppressed.
 We therefore seek another process
that can contribute to rapid cooling.

We exploit the fact that, contrary to a normal electron plasma,
in superconducting proton matter, due to the
Higgs--Meissner effect, the photon acquires
an effective mass that is small compared to the plasmon frequency.
In the region of proton pairing at $T< T_{c,p}$,  we therefore
find that new decay processes of massive photons ($\gamma_m$)
via  electron--electron-hole $(ee^{-1})$ and
 proton--proton-hole ($p\,p^{-1}$)
intermediate states to neutrino--antineutrino pairs,
$\gamma_m\rightarrow e\,e^{-1}+p\,p^{-1}\rightarrow
\nu_{l} \bar{\nu}_{l}$, $l=\{e, \mu, \tau\}$,
can dominate neutron star cooling at certain temperatures.
These processes are determined by the diagrams

\vspace*{-15mm}

\begin{eqnarray}\nonumber
\!\!\!\!\!\!\!\!\!
\setlength{\unitlength}{1mm}
\begin{picture}(50,30)(0,15)
\thicklines
\multiput(2.5,15)(5,0){3}{\line(1,0){2.5}}
\put(25,15){\oval(20,20)}
\put(35,15){\vector(1,1){10}}
\put(35,15){\vector(1,-1){10}}
\put(35,15){\circle*{1.8}}
\put(14,14){\rule{2mm}{2mm}}
\put(25,5){\vector(-1,0){2}}
\put(25,25){\vector(1,0){2}}
\put(24,21){$\displaystyle e$}
\put(24,8){$\displaystyle e^{-1}$}
\put(5,17){$\displaystyle \gamma_m$}
\put(46.5,25){$\displaystyle \nu$}
\put(46.5,5){$\displaystyle \bar\nu$}
\end{picture}
\quad +\,\,\,\,
\setlength{\unitlength}{1mm}
\begin{picture}(50,30)(0,15)
\thicklines
\multiput(0.5,15)(5,0){3}{\line(1,0){2.5}}
\put(25,15){\oval(20,20)}
\put(35,15){\vector(1,1){10}}
\put(35,15){\vector(1,-1){10}}
\put(35,15){\circle*{3}}
\put(13.25,13.25){\rule{3.5mm}{3.5mm}}
\put(25,5){\vector(-1,0){2}}
\put(25,25){\vector(1,0){2}}
\put(5,17){$\displaystyle \gamma_m$}
\put(24,21){$\displaystyle  p$}
\put(24,8){$\displaystyle  p^{-1}$}
\put(46.5,25){$\displaystyle \nu$}
\put(46.5,5){$\displaystyle \bar\nu$}
\end{picture}
\quad +\,\,\,\,
\setlength{\unitlength}{1mm}
\begin{picture}(50,30)(0,15)
\thicklines
\multiput(0.5,15)(5,0){3}{\line(1,0){2.5}}
\put(25,15){\oval(20,20)}
\put(35,15){\vector(1,1){10}}
\put(35,15){\vector(1,-1){10}}
\put(35,15){\circle*{3}}
\put(13.25,13.25){\rule{3.5mm}{3.5mm}}
\put(23.5,5){\vector(1,0){2}}
\put(27.5,5){\vector(-1,0){2}}
\put(23,25){\vector(-1,0){2}}
\put(26,25){\vector(1,0){2}}
\put(5,17){$\displaystyle \gamma_m$}
\put(24,21){$\displaystyle  p$}
\put(24,8){$\displaystyle  p^{-1}$}
\put(46.5,25){$\displaystyle \nu$}
\put(46.5,5){$\displaystyle \bar\nu$}
\end{picture}
\end{eqnarray}

\vspace*{15mm}

\noindent
In the first diagram, the solid lines in the loop are related to Green's
functions of non-superfluid relativistic electrons.
In the second and third diagrams, the solid lines in the loops
correspond to superconducting nonrelativistic protons.
The distinct orientations of arrows indicate that the second
diagram is calculated with so-called ``normal" Green's functions
\setlength{\unitlength}{1mm}
\begin{picture}(10,2)
\thicklines
\put(0,1){\line(1,0){10}}
\put(4,1){\vector(1,0){2}}
\end{picture}, which become the usual Green's functions
for normal Fermi liquids in the limit $\Delta_p \rightarrow 0$.
In contrast,
the third diagram is built up with the ``anomalous" Green's functions
\setlength{\unitlength}{1mm}
\begin{picture}(10,2)
\thicklines
\put(0,1){\line(1,0){10}}
\put(9,1){\vector(1,0){2}}
\put(1,1){\vector(-1,0){2}}
\end{picture}
and
\setlength{\unitlength}{1mm}
\begin{picture}(10,2)
\thicklines
\put(0,1){\line(1,0){10}}
\put(3,1){\vector(1,0){2}}
\put(7,1){\vector(-1,0){2}}
\end{picture}, which are proportional to the proton gap. Therefore
the contribution of the third diagram vanishes for
$\Delta_p\rightarrow 0$. The fat vertices in the second and third
nucleon diagrams include nucleon--nucleon correlations.

The contribution to neutrino production matrix elements of
the third diagram and terms proportional to the gap in the second diagram
is as small as
$(\Delta_p /\epsilon_{Fp})^{2}\ll 1$ for $T<T_{c,p} \ll \epsilon_{Fp}$
(here $\epsilon_{Fp}$ is the proton Fermi energy), compared to
the contribution of the second diagram calculated with
Green's functions of the normal Fermi liquid.
To this same accuracy, we drop the third diagram and
use the Green's functions of protons for the normal Fermi
liquid\footnote{Note that in conventional nuclear
physics one usually employs particle--hole diagrams even at zero
temperature, thereby considering nuclear matter to be normal.
Small effects of pairing can be neglected,
since the typical energy in a nucleonic particle--hole diagram is
of the order of the Fermi energy  $\epsilon_{F}$, and
$\epsilon_{F}\gg \Delta $ holds~\cite{ll,migdal,migrep}.}
in the second diagram.
We thus calculate the emissivity according to
the first two diagrams, assuming $\Delta_{p}=0$ in the
second diagram but taking into account that the photon dispersion relation
is changed due to proton superconductivity.

Our paper is organized as follows.
In Sec. 2 we show that  in the region of proton
superconductivity due to
the Higgs--Meissner effect, the photon spectrum is rearranged, and
instead of the plasmon gap the
photon acquires a mass, which is now determined
by the density of paired protons.
In Secs. 3 and 4 we demonstrate the efficiency of these new processes
in the course of neutron star cooling. The emissivity corresponding
to the above diagrams is calculated and compared
with the emissivity of the standard URCA process and
photon emissivity from the neutron star surface. In Sec. 5 we
detail our conclusions.

\section{Photon spectrum in the superconducting phase}

As is well known \cite{ll}, the photon spectrum in superconducting matter
and in a normal plasma are substantially different.
In the superconducting matter considered here we deal with two subsystems.
The normal subsystem contains
electrons and  non-paired protons and neutrons, which are
present to some extend at finite temperatures. The superfluid
subsystem contains
paired protons and neutrons.
In the presence of a superconducting proton phase,
normal currents associated with both electrons and
residual non-paired protons
are fully compensated by the corresponding response of the
superconducting current \cite{ll,Kin,putt}; otherwise there would be no
superconductivity. What remains after this compensation is a
part of the superconducting current. The resulting
photon spectrum is thereby determined by the inverse of the London
penetration depth ( due to the Higgs--Meissner effect~\cite{ll}),
but not by the plasma frequency, as in the normal system.

In convential superconductors, which contain positively charged ions,
paired electrons, and normal electrons at $T\neq 0$,
the photon spectrum is determined by the relation between the vector
potential $\vec{A}$ and the current $\vec{j}$,
which is proportional to $\vec{A}$;
see Eqs. (96.24) and (97.4) of Ref.~\cite{Kin}. The analogy with the
present case
is straightforward. From the latter equation,
for sufficiently low photon momenta
we immediately obtain the relation
$4\pi \vec{j} \simeq -m_{\gamma}^{2}  (T) \vec{A}$ between the Fourier
components of the current and the vector potential, where
the effective photon mass  is
\begin{equation}  \label{2}
m_{\gamma}  (T)\simeq
\sqrt{\frac{4\pi e^2\rho ^*_p(T)}{m^*_p}}, \quad T < T_{c,p}.
\end{equation}
Here  $m_p^*$ denotes the effective in-medium proton mass,
and $\rho ^*_p(T)=\rho _p(T_{c,p}-T)/T_{c,p}$ denotes the paired proton
density. The choice of a linear temperature dependence for
$\rho ^*_p(T)$ corresponds to the Ginzburg--Landau approach.
A small complex contribution $\sim e^2 f(\omega , k)
\mbox{exp}(-\Delta_p /T)\vec{A}$, where $f(\omega , \vec{k})$ is a function
of the photon frequency $\omega$ and momentum $\vec{k}$,
has been neglected in the above relation between $\vec{j}$ and $\vec{A}$.
More realistically, for $T$ near $T_{c,p}$, one must
take into account this off--shell effect for the photon. At lower
temperatures, correction terms are exponentially suppressed.
Below we take the photon spectrum to be
\begin{equation} \label{photspec}
\omega =\sqrt{{\vec k }^2 + m_\gamma^2},
\end{equation}
thus neglecting the aforementioned small
polarization effects.

Note that external photons cannot penetrate
far into the superconducting region. The photons that we deal
with are thermal
photons with foregoing dispersion law, governed
by the corresponding Bose distribution.
In considering neutrino reactions below, we integrate over the photon
phase-space volume thus accurately  accounting for the distribution of
these photons in warm neutron star matter.

To illustrate more transparently the most
important facets of the reconstruction of the
photon spectrum in the superconducting region, we consider a
two-component,
locally neutral system consisting of
 charged fermions (i.e., the normal subsystem)
described by the Dirac field $\psi$,
and a charged condensate (i.e., the superconducting  subsystem)
described by a condensate wave function
\begin{equation}\label{varphi}
\varphi=\varphi_c\exp{(i\Phi)}.
\end{equation}
The real quantity $\varphi_c$ is the order parameter of the system, i.e.,
$\varphi_c^2 \sim n_c$, where $n_c$  is the number density
of particles in the
condensate,  and the real value $\Phi$ is a phase.
In a fermionic system with pairing, the density $n_c$
is proportional to the pairing gap $\Delta$.

The equation for the electromagnetic field $A_{\mu}$
in such a system reads
\begin{equation}\label{eqA}
\Box A_{\mu}=4\pi\,j_{\mu},
\end{equation}
where the current is
\begin{equation}\label{current}
j_{\mu}=ei\bar \psi\gamma_{\mu}\psi-ei(\varphi^*\partial_{\mu}\varphi-
\varphi\partial_{\mu}\varphi^*)- 2e^2|\varphi|^2A_{\mu}.
\end{equation}
Substituting Eq.~(\ref{varphi}) into
Eq.~(\ref{current}), the electromagnetic current becomes
\begin{equation}\label{current_cond}
j_{\mu}=j_{\mu}^A+\delta j_{\mu},
\end{equation}
where the first term $j_{\mu}^A=-2e^2\varphi_c^2A_{\mu}$
is the superconducting current, and the
second term $\delta j_{\mu}$ contains the normal current $j_{\mu}^{\rm nor}$
and some response $j_{\mu}^{\rm res}$ from the charged condensate, i.e.,
\begin{equation}\label{current_norm}
\delta j_{\mu}=j_{\mu}^{\rm nor}+j_{\mu}^{\rm res}=
ei\bar \psi\gamma_{\mu}\psi+2e\varphi_c^2\partial_{\mu}\Phi_0 .
\end{equation}

Due to gauge invariance, the phase $\Phi = \Phi_0 +\Phi^{\prime}$
is not constrained, and $\Phi_0$
can be chosen in such a way that it cancels the normal current, i.e.,
$\delta j_{\mu}=0$;
otherwise the remaining part of the normal current would destroy
superconductivity and the ground state energy would increase.
This compensation of the normal current $j_{\mu}^{\rm nor}$, which in
metals and in normal plasma
is proportional to the electric field $\vec{E}$,
is a necessary condition for the existence of superconductivity.
Only a diamagnetic part of the fermionic current proportional to the
electromagnetic field $A_{\mu}$ may remain. The latter may lead
only to a minor ($\sim e^2$) contribution
to the unit values of dielectric and diamagnetic constants.
The remaining part of the phase $\Phi^{\prime}$ is hidden in the gauge
field, resulting in the disappearance of the Goldstone field
(see the analogous discussion of the Higgs effect, e.g., in Ref.~\cite{Ber}).
The total number of degrees of freedom does not change, so the
disappearance of the Goldstone field is compensated by the appearance
 of an extra (third) polarization of the photon.
As a result of Eqs.~(\ref{eqA}) and (\ref{current_cond}),
the electromagnetic field obeys the equation
\begin{equation} \label{eqA_cond}
\Box A_{\mu}=-8\pi e^2\varphi_c^2A_{\mu}\,,\end{equation}
which immediately yields the photon spectrum in the form (\ref{photspec}),
where the photon mass is now given by
\begin{equation} \label{photmass}
m_{\gamma} = \sqrt{8 \pi \, e^2 \varphi_c^2}.
\end{equation}

What we have demonstrated is known as the Higgs--Meissner effect:
in the presence of a superconducting component, the photon
acquires finite mass. We see that in a
two-component (normal $+$ superconducting) system, the photon
is described by the dispersion relation (\ref{photspec}),
as it would be in a purely superconducting system, and not by
a plasma-like dispersion law, as in the absence of superconductivity.
Another way to arrive at Eq.~(\ref{photspec}) is given
 in the Appendix in a non-covariant formulation.
Similar derivations for different specific physical systems,
guided by the general
principle of the compensation of the normal currents in a
superconductor, can be found in Refs. \cite{ll,putt,Kin,HS}.

Expressing the amplitude of the condensate field in terms of the
paired proton density~\cite{ll}, one obtains from Eq.~(\ref{photmass})
the result (\ref{2}).
Taking $m_p^*(\rho_0  )\simeq\, 0.8 m_N$
(with $m_N$  the free nucleon mass), with
Eqs.~(\ref{3}) and (\ref{2}) we estimate
$ m_{\gamma}  (\rho =\rho_0  ,T) \simeq \,
1.6 \sqrt{(\displaystyle  T_{c,p}-T)/\displaystyle  T_{c,p}}$~MeV
$\ll \omega _{pl}(\rho  \sim \rho_0)$.
Due to the  rather low effective photon mass
in superconducting neutron star matter at $T<T_{c,p}< \omega _{pl}$, one
may expect a corresponding increase in the contribution of
the above diagrams to  neutrino emissivity.

To avoid misunderstanding, we note the following.
At the first glance one might
suggest that the photon self-energy is completely determined by the above
 neutrino production diagrams, but with
neutrino legs replaced by a photon line. If so,
the contributions of the electron-loop and proton-loop diagrams would
accurately
determine the plasmon spectrum of photon excitations with
energy gap equal to a high plasma frequency
(at least if one drops
small terms proportional to the proton gap  in the
calculation of the proton--proton-hole diagram, now with an incoming and
outgoing photon, as  suggested
for the corresponding neutrino process). How does this relate
to the massive photon spectrum of
superconducting systems? The answer
is that in a system with a charged condensate, in addition to the cited
photon propagation diagrams, there appear
specific diagrams for photon rescattering off the condensate given
by terms $\propto e^2\, \varphi_c^2\, A_{\mu} A^{\mu}$,
$2e\, \varphi_c^2\, \partial_{\mu}\Phi \, A^{\mu}$
in the corresponding Lagrangian.
Their contributions to the equation of motion for the
electromagnetic field are, respectively, the last two condensate terms
in the electromagnetic current in Eq.~(\ref{current}).
The specific condensate diagrams responsible for the
compensation
of the loop diagram contributions in the photon propagator make
no contribution to neutrino
emissivity. Indeed, the neutrino legs cannot be directly
connected to the photon line via such interactions,
(without invoking the internal structure of the condensate
order parameter $\varphi_c$; this contribution is obviously
small compared to what we have taken into account).
Thus, we have argued that in the presence of superconducting protons,
neutrino pairs can be produced in the reaction shown by the
above diagrams, where the photons possess rather small masses generated by the
Higgs--Meissner mechanism.

Having clarified of this important issue,
we are ready to calculate the contribution of these
processes to neutrino emissivity and
compare the result with known emission rates.

\section{Calculation of emissivity}

The matrix element of the above diagrams for the $i$-th neutrino
species ($i=\{\nu_e,\nu_{\mu},\nu_{\tau}\}$) is
\begin{equation} \label{4}
{\cal M}^{(i)a}=-i \sqrt{4\pi} e \frac{G}{2\sqrt{2}} \varepsilon _{\mu}^a
\left(\Gamma_{\gamma}T_p^{(i) \, \mu\rho }-T_e^{(i) \,
\mu\rho }\right) l_{\rho },
\end{equation}
where
\begin{equation}  \label{5}
T_j^{(i) \, \mu\rho } = - \, \mbox{Tr} \int \frac{d^4 p}{(2\pi)^4} \,
\gamma^{\mu} \, i {\hat G}_j(p) \, W^{(i)\rho }_j \, i {\hat G}_j(p+k),
\quad j=\{e,p\},
\end{equation}
and
\begin{equation} \nonumber
{\hat G}_j(p)=
({\hat p} +m_j)\left\{ \frac{1}{p^2-m_j^2} +2\pi \, i \, n_j(p) \,
\delta(p^2-m_j^2)\Theta(p_0)\right\}
\end{equation}
is the in-medium electron (proton) Green's function, and
$n_j(p)=\Theta(p_{Fj}-p)$. $\varepsilon _{\mu}^a$ is the
corresponding polarization four-vector of the massive photon, with
three polarization states in superconducting matter. The factor
$\Gamma_{\gamma}$ takes into account nucleon--nucleon correlations
in the photon vertex. The quantity $G=1.17 \cdot
10^{-5}$~GeV$^{-2}$ is the Fermi constant of the weak interaction.
Above, $l_{\rho }$ denotes the neutrino weak current. The electron
and proton weak currents are
\begin{equation} \label{6}
W_e^{(i)\rho }=\gamma^{\rho }(c_V^{(i)}-c_A^{(i)}\gamma_5), \quad\quad
W_p^{\rho }=\gamma^{\rho }(\kappa_{pp}-g_A\gamma_{pp}\gamma_5),\end{equation}
where $c_V^{(\nu_e)}=c_V^{(+)}=1+4\sin^2\theta_W\simeq 1.92$ and
$c_V^{(\nu_{\mu})}=c_V^{(\nu_{\tau})}=c_V^{(-)}
=1-4\sin^2\theta_W\simeq0.08$. $\theta_W$ is the Weinberg angle
and $c_A^{(\nu_e)}= -c_A^{(\nu_{\mu},\nu_{\tau})}=1$. Proton
coupling is corrected by nucleon--nucleon correlations, i.e., by
the factors $\kappa_{pp}$ and $\gamma_{pp}$ \cite{vs}.

Integrating Eq.~(\ref{5}) over the energy variable, we obtain for
the $i$-th neutrino species
\begin{equation} \label{7}
-i\left(T_p^{(i)\mu\rho }-T_e^{(i)\mu\rho }\right)=
\tau_t^{(i)}\, P^{\mu\rho }+ \tau_l^{(i)}\, F^{\mu\rho }
+\tau_5^{(i)}\, P_5^{\mu\rho },\end{equation}
\begin{equation} \nonumber
P^{\mu\rho }= (g^{\mu\rho }-\frac{k^{\mu}k^{\rho }}{k^2} + F^{\mu\rho }),
\quad
F^{\mu\rho }=\frac{j^{\mu}j^{\rho }}{k^2[(k\cdot u)^2-k^2]},
\quad
P_5^{\mu\rho }=\frac{i}{\sqrt{k^2}}\,
\varepsilon ^{\mu\rho \delta\lambda} k_{\delta}
u_{\lambda},
\end{equation}
where $j^{\mu}=(k\cdot u)k^{\mu}-u^{\mu}k^2$, $k^{\mu}=(\omega
,\vec k )$, $k^2=k_{\mu}k^{\mu}=\omega ^2-\vec k ^2$. The
four-velocity of the medium $u^{\mu}$ is introduced for the sake of
covariant notation.
The transverse ($\tau_t$), longitudinal ($\tau_l$), and axial
($\tau_5$) components of the tensors in Eq.~(\ref{7}) yield
\begin{equation} \label{10}
\tau_t^{(i)}=\tau_{t e}^{(i)}-\tau_{t p}^{(i)}
=2c_V^{(i)}(A_e+k^2B_e)-
2c_V^{(-)} R_{\kappa}(A_p+k^2B_p),\end{equation}
\begin{equation}\nonumber
\tau_l^{(i)}=\tau_{l e}^{(i)}-\tau_{l p}^{(i)}
=4\,k^2[c_V^{(i)}B_e- c_V^{(-)}R_{\kappa}B_p],\end{equation}
\begin{equation}\nonumber
 \tau_5^{(i)}=\tau_{5e}^{(i)}-\tau_{5p}^{(i)}=
(k^2)^{3/2} [c_A^{(i)}C_e-g_A \gamma_{pp} C_p],\end{equation}
where $R_{\kappa}=\kappa_{pp}/c_V^{(-)}$, and
\begin{equation}  \label{11}
A_j=\int\frac{d^3p}{(2\pi)^3}\frac{n_j(p)}{E_p^{(j)}}
+\frac{k^2}{2}\left(1+\frac{k^2}{2m_j^2}\right) m_j C_j,
\end{equation}
\begin{equation} \nonumber
B_j =
\int\frac{d^3p}{(2\pi)^3}\frac{n_j(p)}{2E_p^{(j)}}
\frac{1-\frac{\displaystyle (\vec p \, \vec k )^2}
{\displaystyle  E_p^{(j)2} \vec k ^2}}
{\left(\displaystyle \omega -\frac{\displaystyle \vec p \, \vec k }
{E_p^{(j)}}\right)^2-\frac{\displaystyle  k^4}{\displaystyle  4E_p^{(j)4}}},
\end{equation}\begin{equation} \nonumber
C_j = \int \frac{d^3p}{(2\pi)^3} \, n_j(p)
\frac{m_j}{E_p^{(j)3}}
\left[
\left(\omega -\frac{\vec p \, \vec k }{E_p^j}\right)^2 -
\frac{k^4}{4E_p^{(j)4}}\right]^{-1},
\quad
E_p^{(j)}=\sqrt{m_j^2+ \vec p\,^2}.
\end{equation}
Here we note that the contribution of the axial component $\tau_5$
to the resulting neutrino emissivity
is small ( ${\tau_5}/{\tau_t}\sim m_{\gamma}^2\tau_5/\omega ^2
\tau_l$~$\sim
m_{\gamma}/m_N^*$ for protons and $\sim
(m_{\gamma}m_e/p_{Fe}^2)\ln(p_{Fe}/m_e)$ for electrons), so that it
will be omitted.

The squared matrix element (\ref{4}) for a certain neutrino
species, summed over the lepton spins and averaged over the three
photon polarizations, can be cast in the form
\begin{equation}  \label{13}
{\overline{\sum|{\cal M}^{(i)}|^2}} =
\frac{4}{3}\pi e^2 G^2
\biggl[ \tau_t^{(i)2}\left(2\omega _1\omega _2+
2\frac{(\vec k \vec q _1)(\vec k \vec q _2)}{\vec k ^2}\right)
 \end{equation}\begin{equation} \nonumber
- \tau_l^{(i)2}\left(\omega _1\omega _2+\vec q _1\vec q _2 -2\frac{(k\cdot q_1)(k\cdot
q_2)}{k^2} -2\frac{(\vec k  \vec q _1)(\vec k  \vec q _2)}{\vec k ^2}\right)\Biggr],
\end{equation}
where $(k\cdot q_{1,2})=\omega \omega _{1,2}-(\vec k \vec q
_{1,2})$, and $\omega _{1,2}$ and $\vec q _{1,2}$ denote the
frequencies and momenta of the neutrino and antineutrino. We have
also used the fact that
Tr$\{l^{\mu}l^{\nu}\}=8[q_1^{\mu}q_2^{\nu}+q_2^{\mu}q_1^{\nu}-
g^{\mu\nu}(q_1\cdot q_2)-4i\varepsilon^{\mu\nu\lambda\rho
}q_{1\lambda}q_{2\rho }]$.

The emissivity of our processes is given by
\begin{equation} \label{14}
\varepsilon _{\nu}^{\gamma}=
\int\frac{d^3k}{(2\pi)^3 2\omega } \frac{d^3q_1}{(2\pi)^3 2\omega _1}
\frac{d^3q_2}{(2\pi)^3 2\omega _2}
\frac{\omega _1+\omega _2}{
\displaystyle  \left(e^{\displaystyle \frac{\displaystyle
 \omega _1+\omega _2}{\displaystyle  T}}-1 \right) }
\times \end{equation}\begin{equation} \nonumber
\sum_{i=\nu_e,\nu_{\mu},\nu_{\tau}}
\overline{\sum |{\cal M}^{(i)}|^2}
(2\pi)^4\delta^4(k-q_1-q_2).
\end{equation}
Substituting Eq.~(\ref{13}) into Eq.~(\ref{14}), we finally obtain
\begin{equation} \label{15}
\varepsilon ^{\gamma}_{\nu}=\frac{T^5}{9(2\pi)^3} \pi e^2 G^2   \alpha^2 I,
\qquad
I=\int^{\infty}_{\alpha}\frac{d\xi \xi}{e^{\xi}-1}\sqrt{\xi^2-\alpha^2}
\left[
\tau_{t}^2
(\frac{\alpha^2}{\xi^2})+\tau_{l}^{2}
(\frac{\alpha^2}{\xi^2})\right],\end{equation}
where $\alpha=\frac{\displaystyle  m_{\gamma}  }{\displaystyle  T}$, and
\begin{equation} \label{16}
\tau_t^2(x)
\approx
4 \quad \sum_{i=\nu_e,\nu_{\mu},\nu_{\tau}}
 \left[c_V^{(-)} R_{\kappa}\frac{\rho _p}{2m_p^*}(1+x)
-c_V^{(i)}(\frac{3}{8\pi}\rho _p)^{2/3} (1+\frac{x}{2})
\right]^2,\end{equation}\begin{equation} \nonumber
\tau_l^2(x)
\approx
4x^2  \sum_{i=\nu_e,\nu_{\mu},\nu_{\tau}}
 \left[c_V^{(-)} R_{\kappa}\frac{\rho _p}{2m_p^*}
-c_V^{(i)}(\frac{3}{8\pi}\rho _p)^{2/3} \right]^2.
\end{equation}
Some numerically small terms have been dropped in Eq.~(\ref{16}).

The integral $I$ can be calculated analytically in the two limiting
cases $\alpha\ll 1$ and $\alpha\gg 1$:
\begin{equation} \label{18}
I(\alpha\gg 1)
\approx
\frac{\sqrt{2\pi}}{2} \, \alpha^{3/2} \,
(1+\frac{3}{2\alpha}) \, e^{-\alpha} \,
[{\tau}_{l}^2 (1) + {\tau}_{t}^2
(1)],\end{equation}\begin{equation} \nonumber
I(\alpha\ll 1)
\approx
2 \zeta(3) \, [{\tau}_{l}(0)+
{\tau}_{t}^2 (0)],
\quad\zeta(3)\simeq 1.202.
\end{equation}
Thus, combining Eqs.~(\ref{3}) and (\ref{15})--(\ref{18}), we
obtain an estimate for the emissivity of our reactions (we present
here the result for $ m_{\gamma}   > T$ and for three neutrino
species):
\begin{equation} \label{20}
\varepsilon ^{\gamma}_{\nu}\approx 2.6\cdot10^{25} \, T_9^{\frac{3}{2}} \,
e^{-\frac{\displaystyle  m_{\gamma}  }{\displaystyle  T}}\hspace*{-1.2mm}
\left(
\frac{ m_{\gamma}  }{\mbox{MeV}}
\right)^{\frac{7}{2}}\hspace*{-1.2mm}
\left(\frac{\rho }{\rho_0  }\right)^{\frac{8}{3}} \hspace*{-1.2mm}
\left(1+\frac{3}{2}
\frac{T}{ m_{\gamma}  }\right)
[1+\eta] \,
\frac{\mbox{erg}}
{ {\mbox{cm}}^3\,{\mbox{sec}} } ,
\end{equation}
\begin{equation}  \label{16'}
\eta=0.0003 \, R_{\kappa}^2\left(\frac{m_p}{m_p^*}\right)^2
\left(\frac{\rho }{\rho_0  }\right)^{\frac{4}{3}}
-0.035 \, R_{\kappa}\left(\frac{m_p}{m_p^*}\right)
\left(\frac{\rho }{\rho_0  }\right)^{\frac{2}{3}}.
\end{equation}
Here  $T_9$ denotes  temperature measured in units of
$10^9$~K. The 1 in square brackets in Eq.~(\ref{20}) corresponds to
the electron--electron-hole diagram, whereas the factor $\eta$ is
related to the proton--proton-hole (first term in Eq.~(\ref{16'}))
and the interference diagrams (second term in Eq.~(\ref{16'})).

The emissivity  Eq.~(\ref{20}) varies with temperature as
$T^{3/2}\exp(- m_{\gamma}  /T)$, whereas the emissivity of the
modified URCA process varies as $T^8\exp(-(\Delta_p+\Delta_n)/T)$
in the region of  proton ($\Delta_p\neq 0$) and  neutron
($\Delta_n\neq 0$) pairing. Hence, one can expect that the process
$\gamma_m\rightarrow \nu{\bar\nu}$  will dominate at comparatively
low temperatures, when $\Delta_p(T)+\Delta_n(T)- m_{\gamma}  (T)>0$
and $T<T_{c,p}$.

\section{Numerical estimates}

To obtain quantitative estimates we need the values of the
nucleon--nucleon correlation factors $\kappa_{pp}$  and
$\Gamma_{\gamma}$. According to Ref.~\cite{vs} we can exploit
\begin{equation} \label{21}
\kappa_{pp}=c_V^{(-)}-2f_{np}C_0 A_{nn} \Gamma(f_{nn}),
\end{equation}
where  $f_{np}\simeq -0.75$ and $f_{nn}\simeq 1.25$ are the
constants in the theory of finite Fermi systems \cite{migdal,vs};
$C_0^{-1}=m_n^*p_{Fn}/\pi^2$ is the density of states at the Fermi
surface; $A_{nn}$ is the neutron--neutron-hole loop,
\begin{equation} \label{22}
C_0 \, A_{nn}=iC_0 \int \frac{d^3p}{(2\pi)^4} \, G_n(p+k) \, G_n(p) \,
\approx \frac{p_{Fn}^2 k^2}{6m_n^*\omega ^2},
\end{equation}
for values of $\omega \gg |\vec k |p_{Fn}/m_n^*$ of interest, and
$\Gamma^{-1}(f_{nn})=1-2f_{nn}C_0A_{nn}$.

We note that the second term in Eq.~(\ref{21}) is not proportional
to a small factor $c_V^{(-)}$, because the nucleon--nucleon
correlations also allow for the emission of $\nu\bar\nu$-pairs from
the $nn^{-1}$ loop. Numerical estimates of the ratio $R_{\kappa}$
are as follows: for $\alpha\gg 1$, $R_{\kappa}\simeq 1.6$ for $\rho
=\rho_0  $, $m_n^*(\rho_0  )\simeq 0.8 m_n$, and $R_{\kappa}\simeq
2.1$ for $\rho =2\rho_0  $, $m_n^*(2\rho_0  )
\simeq 0.7
m_n$; for $\alpha\ll 1$, $R_{\kappa} \simeq 1$ and correlation
effects are negligible. The in-medium renormalization of the proton
electric charge included in the factor $\Gamma_{\gamma}$ can be
also expressed in terms of the constants in the theory of finite
Fermi systems and the proton--proton loop factor ($A_{pp}$); see
Ref.~\cite{migdal}. The latter is suppressed at relatively low
proton densities. We can therefore take $\Gamma_{\gamma}\approx 1$.
With these estimates, we observe that the main contribution to
neutrino emissivity comes from electron--electron-hole processes.

The ratio of the emissivity $\varepsilon ^{\gamma}_{\nu}$
(\ref{20}) to the emissivity $\varepsilon _{\nu}^{FM}$ of the
modified URCA process, $R_{FM}  =  {\varepsilon
^{\gamma}_{\nu}}/{\varepsilon _{\nu}^{FM}}$, is
\begin{equation}\label{24}
R_{FM}
 \approx
1.5 \cdot 10^{4} \cdot T_9^{-13/2}  \,
e^{\frac{\Delta_n+\Delta_p- m_{\gamma}  }{ T}}
\left(\frac{ m_{\gamma}  }{\mbox{MeV}}\right)^{\frac{7}{2}}
\left(1+\frac{3}{2}\frac{T}{ m_{\gamma}  }\right)
\left(\frac{\rho }{\rho_0  }\right)^2
\left(\frac{m_n^3 \, m_p}{m_n^{*3} \, m_p^*} \right)
[1 + \eta].
\end{equation}

For further estimates we need the values of the neutron and proton
gaps, which are unfortunately model--dependent. For instance, the
evaluation in Ref.~\cite{takatsuka} yields $\Delta_n(0)
\simeq\,8.4\, T_{c,n} \simeq\, 0.6$~MeV, $T_{c,n}\simeq 0.07$~MeV
for $3P_2$ neutron pairing at $\rho =\rho_0  $, and $\Delta_p(0)
\simeq\,1.76\, T_{c,p} \simeq\, 3$~MeV, $T_{c,p}\simeq 1.7$~MeV for
$1S$ proton pairing, while Ref.~\cite{pines} uses $\Delta_n(0)
\simeq\,2.1$~MeV, $T_{c,n}\simeq 0.25$~MeV  and $\Delta_p(0)
\simeq\,0.7$~MeV, $T_{c,p}\simeq 0.4$~MeV for $\rho =\rho_0  $.
Employing these estimates of the zero-temperature gaps, their
temperature dependence, and the photon effective mass, we obtain
from Eq.~(\ref{24}) the temperature dependence of the ratio
$R_{FM}$.

In order to find the lower temperature limit at which the processes
$\gamma_m\rightarrow \nu\bar\nu$ are still operative, we need to
compare the value $\varepsilon ^{\gamma}_{\nu}$ with the photon
emissivity at the neutron star surface, $\varepsilon
_{\gamma}^s=3\sigma T_s^4/R$, where $\sigma$ is the
Stefan--Boltzmann constant, $T_s$ denotes the  surface temperature
of the star, and $R$ is the star's radius. By employing a relation
\cite{gud} between the surface and interior temperatures, we obtain
for $R_{\gamma}= {\varepsilon
_{\nu}^{\gamma}}/{\varepsilon _{\gamma}^s}$
\begin{equation} \label{25}
R_{\gamma}
\approx\, 1.2\cdot 10^9 T_9^{-0.7} e^{-\frac{\displaystyle  m_{\gamma}  }
{\displaystyle  T}}
\left(\frac{ m_{\gamma}  }{\mbox{MeV}}\right)^{\frac{7}{2}}
\left(1+\frac{3}{2}\frac{T}{ m_{\gamma}  }\right)
\left(\frac{\bar{\rho }}{\rho_0  }\right)^{\frac{8}{3}}[1+\eta],
\end{equation}
where the star's radius and mass are taken to be $10$~km and $1.4
M_{\odot}$, with $M_{\odot}$ the solar mass and $\bar{\rho}$ some
averaged value of the density in the neutron star interior.

The ratios $R_{FM}$ and $R_{\gamma}$ are plotted as a function of
the temperature in Fig.~1 for both of the foregoing parameter
choices. We see that our new processes are operative in the
temperature range $1\cdot 10^9$~K$\stackrel{\scriptstyle
<}{\phantom{}_{\sim}} T\stackrel{\scriptstyle <}{\phantom{}_{\sim}}
8\cdot 10^9$~K for the parameter choice of Ref.~\cite{takatsuka},
and $1\cdot 10^9$~K$
\stackrel{\scriptstyle <}{\phantom{}_{\sim}}
T\stackrel{\scriptstyle <}{\phantom{}_{\sim}} 4\cdot 10^9$~K for
the parameters of Ref.~\cite{pines}. As one observes in Fig.~1,
within these intervals the new cooling channel might exceed known
cooling processes by up to a factor $10^6$.

\section{Concluding remarks}
 As mentioned above, for $T>T_{c,n}, T_{c,p}$, i.e., in a normal
plasma region of the star's crust and interior, photons with
approximately the electron plasma frequency\footnote{A rather small
extra contribution also comes from the proton--proton-hole
diagram.} $\omega
_{\rm pl}$ can decay into neutrino pairs, as has been shown in
previous estimates~\cite{arw}. At $T<T_{c,p}$, however, we
are already dealing with massive photons in the region of proton
pairing, and our new reaction channels can significantly contribute
to cooling.

 Our processes can also occur in a charged-pion (or kaon)
condensate state but they are suppressed due to the high effective
photon mass\footnote{For simplicity, in this estimate the
peculiarities of a condensate with nonvanishing momentum
\cite{migrep} are ignored.} $ m_{\gamma}
\simeq\sqrt{8\pi e^2 \varphi^2_c}\simeq 6$~MeV for the condensate
field $\varphi_c\simeq 0.1 m_{\pi}\simeq 14$ MeV.

 In deriving the value of $\varepsilon _{\nu}^{FM}$ used above, one
describes the nucleon--nucleon interaction essentially by free
one-pion exchange. In reality, however, at $\rho > (0.5 - 1)\rho_0$
the total nucleon--nucleon interaction does not reduce to free
one-pion exchange, because of the strong polarization of the
medium, whereby a significant part comes from in-medium pionic
excitations \cite{migrep,vs,pion,sv}. Occurring in intermediate
states of the reaction, the in-medium pions can also decay into
$e\bar\nu$, or first into a nucleon--nucleon-hole, which then
radiates $e\bar\nu$, thereby substantially increasing the resulting
emissivity. Other reaction channels such as $n\rightarrow
n_{pair}\nu\bar\nu$ and $p\rightarrow p_{pair}\nu\bar\nu$ open up
in the superfluid phase with paired nucleons~\cite{flowers,vs,sv}. All
these reaction channels give rise to a larger contribution to the
emissivity than that of the modified URCA process estimated via
free one-pion exchange. Above we compared $\varepsilon
_{\nu}^{\gamma}$ with $\varepsilon _{\nu}^{FM}$ just
because the latter is used in the standard scenarios of neutron
star cooling.

As we also mentioned in the Introduction, there are other processes
like those considered above. Emissivity of  the process $p\gamma_m
\rightarrow p_{pair}\nu \bar{\nu}$ is substantially suppressed (at
least by a factor $e^2$ and also due to a much smaller phase-space
volume) compared to that of the process $p\rightarrow
p_{pair}\nu\bar\nu$. According to simple estimates, e.g., using Eq.
(22) of Ref.~\cite{Rit}, the process $e\gamma \rightarrow e \nu
\bar{\nu}$ makes a very small contribution to the emissivity both
in the inner crust and in the interior of neutron stars, even when
one neglects the photon mass. Thus we may conclude that the process
$e\gamma_m \rightarrow e \nu \bar{\nu}$ also leads to a minor
contribution to the emissivity at the densities and temperatures
under consideration.

In summary, the processes $\gamma_m\rightarrow e\,e^{-1} +
p\,p^{-1}\rightarrow \nu\bar\nu$ might be operative over some
temperature interval $T\simeq (10^9-10^{10})$~K, $T<T_{c,p}$, and
together with other in-medium modified processes~\cite{SVSWW}, they
should be incorporated into computer simulations of neutron star
cooling.\\[3mm]

We acknowledge V.\,M. Osadchiev for fruitful discussions. The
research described in this publication was made possible in part by
Grants N3W000 from the International Science Foundation and N3W300
from the International Science Foundation and the Russian
Government. B.\,K. and E.\,E.\,K. are supported by  BMBF Grant
06DR666. E.\,E.\,K. acknowledges the support of the
Heisenberg--Landau program.

\section*{Appendix}

We can also achieve  the same results that led to
Eq.~(\ref{photmass}) by starting with Maxwell's equations (in
obvious notation):
\begin{equation}
i\vec{k} \cdot \vec{E}=4\pi \tilde\rho,\quad i\vec{k} \times
\vec{B}=4\pi \vec{j} -i\omega  \vec{E},
\end{equation}
\begin{equation}
\vec{k}\cdot \vec{B}=0,
\quad
\vec{k} \times\vec{E}=\omega\vec{B},
\end{equation}
where the charge density $\tilde\rho$ is the superposition of the
density of free charges and the density of bound charge. Full free
charge density being zero in our case due to local
electro-neutrality. The current $ \vec j$ is a superposition of an
external test current and the induced current
$$\vec{j}=\vec{j}^{\rm ext}+\vec{j}^{\rm ind}.$$
In normal systems, the induced current (i.e., the current of
non-paired charged particles) $\vec{j}^{\rm ind}=\vec j^{\rm nor}$
is related to $\vec E$ via longitudinal and transverse dielectric
constants $\epsilon_l$ and $\epsilon_t$. This connection  results
in longitudinal and transverse branches of the electromagnetic
excitations, with an effective photon gap equal to the plasma
frequency $\omega_{pl}$ \cite{arw}. In contrast, in a
superconducting system the condensate makes two other contributions
to the current, namely $\vec{j}^{A}=-2e^2
\varphi_c^2\vec{A}$ and $\vec{j}^{\rm res}=2e\varphi_c^2\nabla
\Phi$. Letting $\Phi =\Phi_{(1)}+\Phi_{(2)}$, we have
$\vec{j}^{\rm res}=\vec{j}^{\rm res}_{(1)}+\vec{j}^{\rm
res}_{(2)}$. These two terms are determined as follows. As we have
argued above, superconductivity requires the compensation of the
normal component of the current proportional to $\vec{E}$, i.e., we
can take $\vec{j}^{\rm nor}+\vec{j}^{\rm res}_{(1)}$ = 0. Only
small contributions $\sim e^{2}\mbox{exp}(-\Delta_p /T)
\omega^2 \vec{A}$ and $\sim e^2 \mbox{exp}(-\Delta_p /T)\vec{k}^2 \vec{A}$,
as well as a small imaginary contribution $\sim i e^2 F(\omega ,
\vec{k}) \mbox{exp}(-\Delta_p /T)\vec{A}$, where $F$ is some
function of $\omega$ and $\vec{k}$, can still remain from the value
$\vec{j}^{nor}$ (see Eqs (96.24) and (97.4) of Ref.~\cite{Kin}). We
neglect these small contributions. The part of the current $\sim
\nabla \Phi_{(2)}$ can be hidden in $\vec{j}^{A}$ by a
gauge transformation of the field $\vec{A}$. We then have
$$i\vec{k} \times \vec{B}\simeq \vec{j}^{A} -i\omega  \vec{E}.$$
Taking the vector product of this equation  with $\vec{k}$, we
obtain $(\omega ^2-\vec{k}^2 -8\pi e^2\varphi_c^2)\vec{B} = 0$.
From this relation we observe that the electromagnetic excitations
possess the mass  given by Eq.~(\ref{photmass}). Hence, we have
demonstrated that one can obtain the well-known plasma photon
spectrum for a normal system, and at the same time one can obtain a
massive photon spectrum and the Higgs--Meissner effect in a system
with a charged condensate.

\newpage

\begin{figure}[h]
\begin{center}
\psfig{file=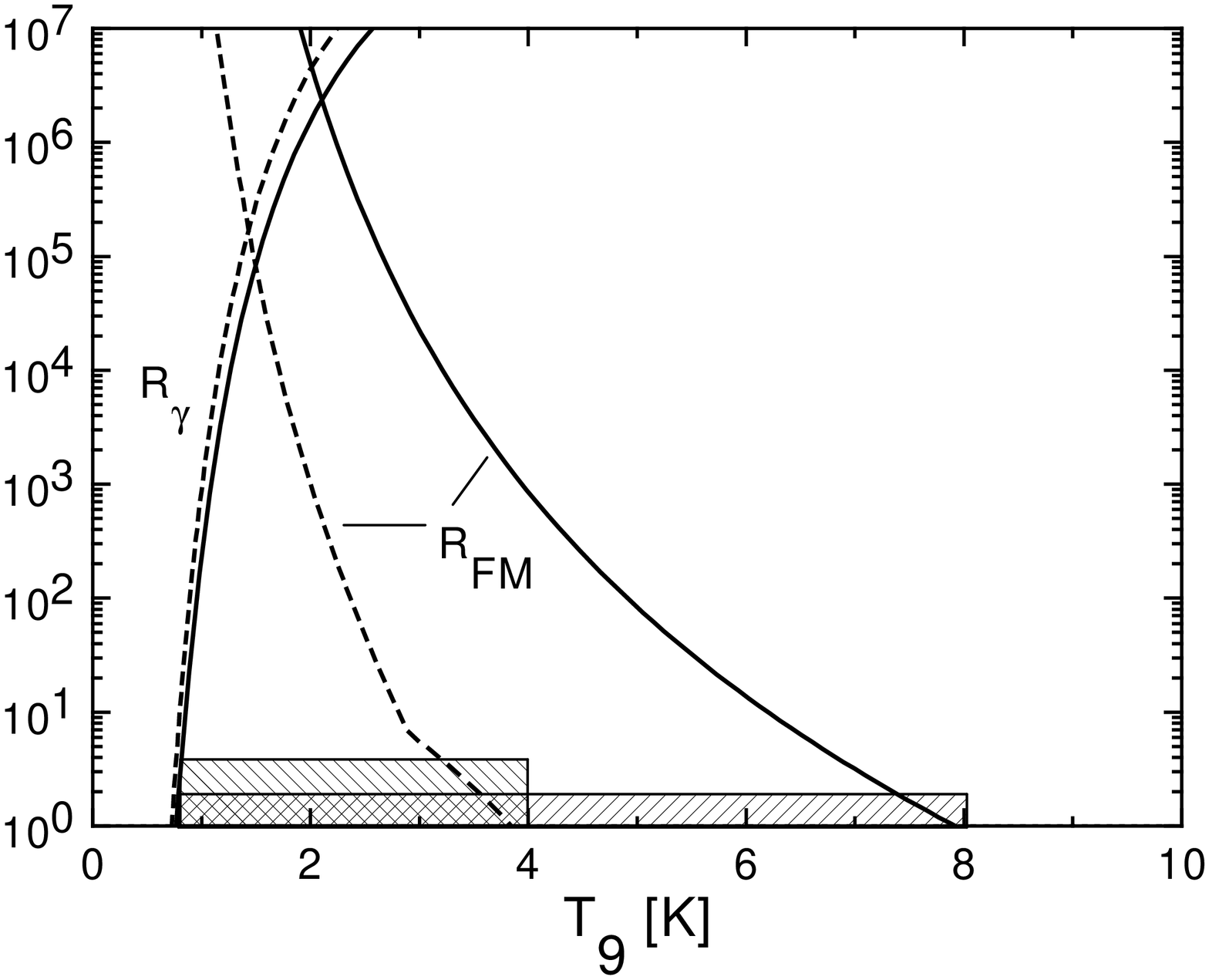,width=15cm,angle=-90}
\end{center}
\caption{
Temperature dependence of the ratios $R_{FM}$ and
$R_{\gamma}$ at nucleon density $\rho =\rho_0  $. Solid curves
correspond to
 the parameter choice of
Ref.~\protect\cite{takatsuka}, whereas the dashed curves depict
results with parameters of ref.~\protect\cite{pines}. Shaded bars
indicate the temperature regions in which cooling via massive
photon decay is more efficient  than standard cooling processes.
}
\end{figure}
\end{document}